\documentclass[epj,nopacs]{svjour}
\usepackage{amsmath}
\usepackage{amssymb}
\usepackage{latexsym}
\usepackage[dvips]{graphicx}
\usepackage[english]{babel}

\def\bge{\begin{equation}}
\def\ene{\end{equation}}
\def\bgea{\begin{eqnarray}}
\def\enea{\end{eqnarray}}

\def\bge{\begin{equation}}
\def\ene{\end{equation}}
\def\bgea{\begin{eqnarray}}
\def\enea{\end{eqnarray}}

\def\ls{\raise 1.5pt\hbox{$\,<\;$}\kern -10.5pt\lower3.5pt
          \hbox{$\sim$}\kern 1.5pt} 
\def\gs{\raise 1.5pt\hbox{$\,>\,$}\kern -9.5pt\lower3.5pt
          \hbox{$\sim$}\kern 1.5pt} 

\usepackage{color}

\usepackage{ulem} 

\begin{document}
\sloppy
\title{The cosmic timeline implied by the highest redshift quasars}
\author{Fulvio Melia\thanks{John Woodruff Simpson Fellow.}}
\institute{Department of Physics, the Applied Math Program, and Department of Astronomy, \\
              The University of Arizona, Tucson, AZ 85721,
              \email{fmelia@email.arizona.edu}}

\authorrunning{Melia}
\titlerunning{{\it JWST} Observations of the Early Universe}

\date{\today}

\abstract{The conventional picture of supermassive black-hole growth in the standard model had
        already been seriously challenged by the emergence of $\sim 10^9\;M_\odot$ quasars
        at $z\sim 7.5$, conflicting with the predicted formation of structure in the early
        $\Lambda$CDM Universe. But the most recent {\it JWST} discovery of a
	$\sim 10^8\;M_\odot$ source at $z\sim 10.1$ argues even more strongly
	against the possibility that these black holes were created in Pop II or III supernovae, 
	followed by Eddington-limited accretion. Attempts at resolving this anomaly have largely
        focused on the formation of seeds via an exotic, direct collapse of primordial
        gas to an initial mass $\sim 10^5\;M_\odot$---a process that has never been seen
        anywhere in the cosmos. Our goal in this {\it Letter} is to demonstrate that the
        emergence of these black holes is instead fully consistent with standard astrophysics
        in the context of the alternative Friedmann-Lema\^itre-Robertson-Walker cosmology
        known as the $R_{\rm h}=ct$ universe. We show that, while the predicted evolution
        in the standard model is overly compressed, the creation, growth and appearance of
        such high-$z$ quasars fall comfortably within the evolutionary history in this
        cosmology, thereby adding considerable observational support to the existing body
        of evidence favoring it over the standard scenario.}
\maketitle

\section{Introduction}\label{Introduction}
The recent discovery \cite{Bogdan:2023} of an X-ray luminous supermassive black hole, 
UHZ-1, at a confirmed spectroscopic redshift $z=10.073\pm 002$ \cite{Goulding:2023},
emphasizes more than ever the time compression problem in the early $\Lambda$CDM Universe
\cite{Melia:2020}. With an inferred mass of $M=10^7-10^8\;M_\odot$, this object should 
have taken over $\sim 700$ Myr to grow via standard Eddington-limited accretion, starting 
with a supernova remnant mass of $\sim 10\;M_\odot$ (see Eq.~\ref{eq:Salpeter} below). 
Yet it appears to us a 
mere $\sim 300$ Myr after Pop III stars started forming some $\sim 200$ Myr beyond the 
big bang. On the flip side, its host galaxy is estimated to have a stellar mass
$M_*\sim 1.4^{+0.3}_{-0.4}\times 10^8\;M_\odot$ \cite{Goulding:2023}, so the
ratio $M_*/M$ is two to three orders of magnitude smaller than local values, consistent
with a scenario in which the black hole formed first within the core of an isolated
Pop III remnant \cite{Haiman:1996,Tegmark:1997,Bromm:2002}, followed by the
gradual aggregation of proto-galactic stars around it. The main problem therefore
seems to be the timeline in $\Lambda$CDM.

\section{Background}\label{background}
The growth of black-hole seeds (regardless of their mass) in conventional
astrophysics is constrained by the maximum accretion rate allowed by the
outward radiation pressure associated with the luminosity produced via
the dissipation of gravitational energy \cite{Melia:2009}. When the ionized
plasma is hydrogen-rich, this limit is known as the Eddington value,
$L_{\rm Edd}\approx 1.3\times 10^{38}(M/M_\odot)$ ergs s$^{-1}$. The unknown
factor in this process is the efficiency, $\epsilon$, with which rest-mass
energy is converted into radiation, fixing the accretion rate,
$\dot{M}=L_{\rm bol}/\epsilon c^2$, in terms of the actual bolometric
luminosity $L_{\rm bol}$, which may be different from $L_{\rm Edd}$. Taking
all possible variations of accretion-disk theory into account, one generally
adopts a fiducial value $\epsilon=0.1$ for this quantity \cite{Melia:2009}.

Thus, if the accretion rate is Eddington-limited, one may combine these
simple expressions to derive the black-hole growth rate,
\begin{equation}
{dM\over dt}={1.3\times 10^{38}\;{\rm ergs/s}\over\epsilon c^2M_\odot}\;M\label{eq:dMdt}
\end{equation}
\cite{Salpeter:1964,Melia:2013a}.
Its straightforward solution is known as the Salpeter relation,
\begin{equation}
M(t) = M_{\rm seed}\exp\left({t-t_{\rm seed}\over 45\;{\rm Myr}}\right),\label{eq:Salpeter}
\end{equation}
where $M_{\rm seed}$ ($\sim 5-25\;M_\odot$) is the supernova-remnant seed mass
created at time $t_{\rm seed}$.

\begin{figure*}
\vskip 0.1in
\centering
\includegraphics[width=0.9\linewidth]{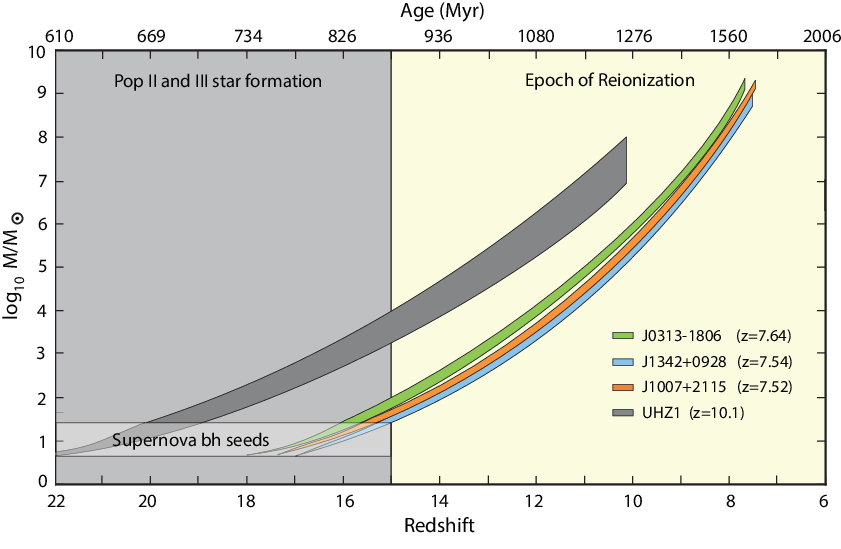}
\caption{Growth history in the $R_{\rm h}=ct$ cosmology of the four
most distant quasars discovered thus far. Pop II and III supernova
seeds formed with initial masses $5\;M_\odot\lesssim M_{\rm init}
\lesssim 25\;M_\odot$ at redshifts $22\gtrsim z\gtrsim 15$, and
grew via Eddington-limited accretion to a mass $M\sim 10^7-10^8\;
M_\odot$ in one case, and $M\sim 10^9\;M_\odot$ by redshift
$\sim 7.5$ for the other three. This seed formation corresponds very
well with our understanding of how the earliest stars formed (and died),
promoting the transition from the `dark ages' to the Epoch of
Reionization (EoR) at $z\sim 15$. The width of each swath indicates
the possible range of $M$ versus $z$ given the unknown seed mass and the
error in $M$ at $z=10.1$ and $\sim 7.5$. If the initial seed were
$M_{\rm init}\sim 25\;M_\odot$, however, all four of these supermassive
black holes would have begun their growth at $20\lesssim z\lesssim 15$,
just prior to the onset of the EoR.}\label{fig1}
\end{figure*}

\begin{figure*}
\vskip 0.1in
\centering
\includegraphics[width=0.9\linewidth]{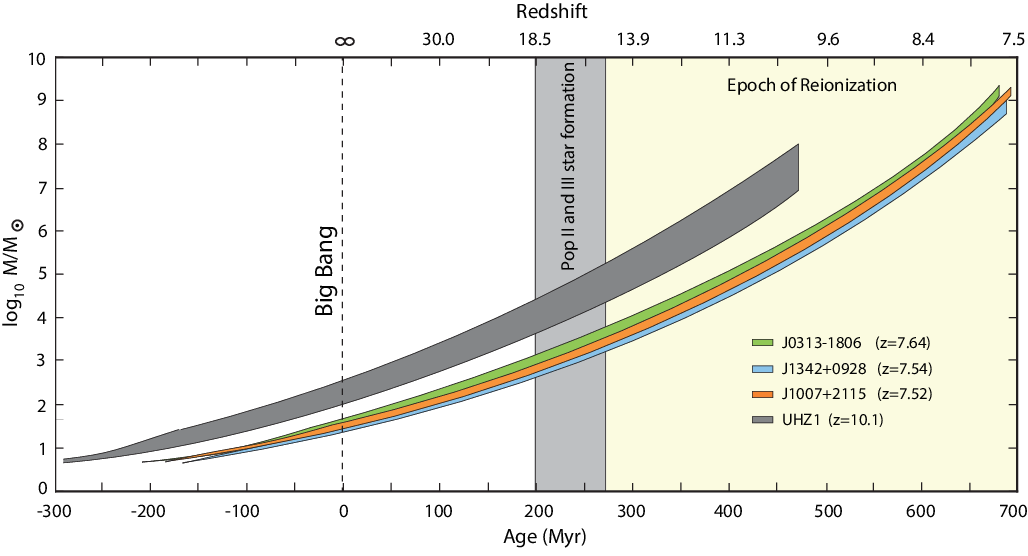}
\caption{Growth history of the four most distant quasars in 
{\it Planck}-$\Lambda$CDM. The seeds have the same mass as those
in figure~\ref{fig1}, but here their formation had to take
place well before the Big Bang. In this case, the birth and growth
of such supermassive black holes do not comport at all with
our understanding of how the earliest stars formed (and died),
and the transition from the `dark ages' to the Epoch of
Reionization (EoR) at $z\sim 15$.}\label{fig2}
\end{figure*}

What is particularly challenging to the standard model is that there is no
evidence of UHZ-1 accreting at a greatly super-Eddington rate, which would be required
to account for its anomalously rapid growth within such a short time
\cite{Volonteri:2005,Inayoshi:2016}. This follows a well-defined pattern
in which the inferred luminosity in other high-$z$ supermassive black holes with
reasonably estimated masses has thus far been at, or near, the Eddington value
(see, e.g., fig.~5 in \cite{Willott:2010a}). Specifically for the bolometric
luminosity of the other three quasars we shall be discussing in this {\it Letter},
J0313-1806 ($z=7.642$) is accreting at $0.67\pm0.14\;L_{\rm Edd}$ \cite{Wang:2021b},
J1342+0928 ($z=7.54$) at $1.5^{+0.5}_{-0.4}\;L_{\rm Edd}$ \cite{Banados:2018},
and J1007+2115 ($z=7.515$) at $1.06\pm0.2\;L_{\rm Edd}$ \cite{Yang:2020}. Quite
alarmingly, all four of these sources, but particularly UHZ-1, would thus have had
to start growing {\it before} the big bang, which is unrealistic
\cite{Melia:2013a,MeliaMcClintock:2015}.

Some attention has thus been given to the massive seed proposal
\cite{Latif:2013,Alexander:2014}, which would necessitate
the birth of black holes with a mass $\sim 10^5\;M_\odot$ at $z\sim 30$,
even before the formation of Pop II and III stars. But this scenario
is even more difficult to confirm observationally. The catastrophic
events creating such massive objects would likely be too brief for
us to see them directly. Observationally, these `intermediate-mass'
black holes could be detected after they formed nearby, but the evidence
is sparse and inconclusive. For example, the low-luminosity active galactic
nucleus NGC~4395 at 4 Mpc may be harboring a $\sim 3.6\times 10^5\;M_\odot$
black hole in its core \cite{Peterson:2005}. Perhaps ultra-luminous X-ray
sources in neighboring galaxies may be black holes with $M\sim 1,000\;M_\odot$
\cite{Maccarone:2007}, though this is well below what is required. It is
also possible that intermediate-mass black holes may have been discovered
in globular clusters, but none has yet stood up to careful scrutiny
\cite{Baumgardt:2003}. The conclusion seems to be that the creation of
massive seeds may be contemplated theoretically, but none has yet been
found. And anyway, if some are discovered in dwarf active galactic nuclei,
they may very well have simply grown to their observed intermediate mass
via steady accretion rather than having been created by some exotic event.

\section{Blackhole growth in $R_{\rm h}=ct$}\label{growth}
In this {\it Letter}, we shall demonstrate that, in contrast to the major
difficulties one faces in accounting for the `too-early' appearance of supermassive
black holes in the context of $\Lambda$CDM, all of the characteristics of these
objects, including their mass and redshift, and our current understanding of
how and when the first stars formed and then died as Pop III supernovae, are
remarkably consistent with the time versus redshift relation in the
$R_{\rm h}=ct$ cosmology (see figure~\ref{fig1}). This issue has been addressed
before with the growing number of quasars at redshifts $z\gtrsim 7$
\cite{Melia:2013b,MeliaMcClintock:2015,Fatuzzo:2017}. In all these cases,
the time compression problem faced by $\Lambda$CDM has been 
eliminated by the use of $R_{\rm h}=ct$ as the background cosmology.

But the most recent discovery of UHZ-1 even closer to the big bang has raised
the tension considerably. Nevertheless, we shall show that all of these
quasars---those at $z\gtrsim 7.5$ and this most recent addition to the
earliest supermassive black-hole family---are self-consistently accounted
for by the formation of supernova remnant seeds at $22\lesssim z\lesssim 15$,
and their subsequent Eddington-limited growth through the Epoch of Reionization
(EoR) beginning at $z\sim 15$, to the redshift at which they were discovered.

The pivotal events believed to have occurred in the early Universe may be briefly
described as follows, based on many detailed simulations carried out in recent years
\cite{Barkana:2001,Miralda:2003,Bromm:2004,Ciardi:2005,Glover:2005,Greif:2007,Wise:2008,Salvaterra:2011,Greif:2012,Jaacks:2012,Bromm:2009,Yoshida:2012}.
At least in the context
of $\Lambda$CDM, with concordance parameters $\Omega_{\rm m}=0.307$,
$k=0$, $w_\Lambda=-1$ and Hubble constant $H_0=67.7$ km s$^{-1}$ Mpc$^{-1}$
\cite{Planck:2016}, the Universe became transparent at $t\sim 0.4$ Myr,
heralding the beginning of the so-called Dark Ages lasting until the
first (Pop III) stars formed $\sim 200-300$ Myr later, at the core of
mini halos with mass $\sim 10^6\;M_\odot$
\cite{Haiman:1996,Tegmark:1997,Bromm:2002}.

There is still some debate about whether this delay of $\sim 200$ Myr between
the big bang and the creation of the first stars could be circumvented somehow,
to mitigate the time compression problem in $\Lambda$CDM, but it is difficult
to see how the primordial gas could have cooled any faster.  By comparison,
this time corresponded to $z\sim 70$ in the $R_{\rm h}=ct$, as we shall see
shortly via the use of Equation~(\ref{eq:redtime}).

A further delay of $\gtrsim 100$ Myr \cite{Yoshida:2004,Johnson:2007}
would have ensued while the hot gas expelled by Pop III stars cooled
and re-collapsed, facilitating the formation of Pop II stars. All told,
the supernova remnant seeds growing into supermassive black holes
could not have formed earlier than $\sim 300$ Myr after the big bang.
Once these stars, and the black holes they subsequently spawned,
started emitting UV radiation, the Universe initiated a transition to
reioniozation, a process that observationally lasted over the redshift
range $15\lesssim z \lesssim 6$ \cite{Zaroubi:2013,Jiang:2006}. In the
standard model, the EoR therefore stretched over a cosmic time
$400\lesssim t\lesssim 900$ Myr, in significant tension with the
timeline required for the supermassive black holes to grow to their
observed mass.

These essential features of black-hole growth in $\Lambda$CDM
are shown in figure~\ref{fig2}, along with the corresponding evolutionary
trajectories in this model of the four most distant quasars highlighted
in figure~\ref{fig1}. The discordance between the theoretical predictions 
and the observational constraints is quite evident in this case, showing
that our current understanding of how structure formed in the early
Universe is inconsistent with the appearance of such massive objects
so quickly after the Big Bang. Indeed, if the seeds for these objects
were $\sim 5-25\;M_\odot$ black holes, and if they subsequently grew
at about the Eddington rate, they would have had to start well before
the Big Bang itself, which is clearly unphysical.

In the $R_{\rm h}=ct$ universe, however, the redshift-time relation
is given by the expression
\begin{equation}
1+z = {t_0\over t}\;,\label{eq:redtime}
\end{equation}
where $t_0=H_0^{-1}$ is the age of the Universe today
\cite{MeliaShevchuk:2012,Melia:2020}. Thus, for the same Hubble
constant $H_0$, the Dark Ages in this model ended at $\sim 878$ Myr,
while the EoR extended from $\sim 878$ to $\sim 2$ Gyr (see fig.~\ref{fig1}).
It is important to emphasize that the EoR redshift range is measured
observationally, independently of the cosmology. What differs between
the two models, however, is the mapping of redshift to age.

Thus, with $R_{\rm h}=ct$ as the background cosmology,
J1007+2115 ($z=7.515$) is being viewed at cosmic time
$t\sim 1.65$ Gyr, about $770$ Myr after the EoR began.
J1342+0928 ($z=7.54$) is being viewed at $\sim 1.64$ Gyr,
and J0313-1806 ($z=7.642$) at $\sim 1.62$ Gyr. According to
the Salpeter relation (Eq.~\ref{eq:Salpeter}), they were
created at $t\sim 733-765$ Myr if their seeds were $\sim 10\;M_\odot$,
or $\sim 829-849$ Myr if the seeds were instead $\sim 25\;M_\odot$.
The redshift range of their formation would therefore have been
$18\gtrsim z \gtrsim 15$, ideally placed just prior to the
onset of the EoR, where one would expect the peak in Pop II and
III star formation---and their supernova deaths---to have occurred.

But the timeline disparity between $\Lambda$CDM and the early quasars
could not be greater than that exhibited by UHZ-1. Detailed studies
\cite{Alvarez:2009,Kim:2011a}, including the state-of-the-art Renaissance
simulation suite \cite{Smith:2018}, have shown that the kind of greatly
super-Eddington accretion required to produce an object like UHZ-1
in only $\sim 300$ Myr is probably infeasible given the spatial location
of the Pop III supernova seeds.

In $R_{\rm h}=ct$, we are viewing UHZ-1 at cosmic time $t\sim 1242$ Myr.
According to the Salpeter equation, it would have taken this object
$\sim 750$ Myr to grow from a $10\;M_\odot$ seed, or $\sim 580$ Myr
from $25\;M_\odot$. The implied range of redshifts for its birth is
thus $27\gtrsim z\gtrsim 20$, as one may see in figure~\ref{fig1}.
Like the other three high-$z$ quasars shown in this figure, its
origin and growth are thus consistent with our current
understanding of Pop III star formation and the subsequent accretion
these objects would have experienced during their lifetime up to the
point at which we detect them.

We must also stress that the $\sim 580-750$ Myr required for UHZ-1 to
grow via Eddington-limited accretion from its initial Pop III supernova
seed at $z\gtrsim 20$ coincides beautifully with two critically
important observations. First it was created not long before the onset
of the EoR, which is believed to have been sustained by UV photons
emitted by the first stars and the black holes they spawned. Second,
there is no evidence that it is accreting at a super-Eddington rate.
The latter is commonly observed in all the high-$z$ quasars, arguing
against anomalously rapid growth as the reason for their appearance
much earlier than expected in the standard model.

Of course, it may be too early to claim that the timeline
required by the `too-early' appearance of supermassive black holes
is uniquely a feature of just $R_{\rm h}=ct$. Other possibilities
exist for stretching the time versus redshift relation predicted
by the standard model. For example, if for some reason $\Omega_{\rm m}$
were to be $0.1$ instead of the value $0.307$ optimized by {\it Planck},
then $t(z=10)\sim 0.8$ Gyrs in $\Lambda$CDM, which would be consistent 
with the features of black-hole growth we have described in this 
{\it Letter}. Alternatively, one might contemplate the presence of
dynamical dark energy instead of a cosmological constant \cite{Zhao:2017},
which could also impact the timeline in the early Universe. It should
be mentioned, however, that adjusting the timeline cannot be done 
independently of all the other observational constraints. In the case 
of $R_{\rm h}=ct$, this timeline is built in from the beginning and 
is consistent with all the $\sim 30$ comparative tests completed thus 
far \cite{Melia:2020}. It would be more challenging to ensure that
modifications such as $\Omega_{\rm m}\rightarrow 0.1$ would retain
global consistency of the standard model with all of the data
available today.

In a very different scenario, the seeds for supermassive black-hole
growth might have been primordial black holes, possibly formed around 
the electroweak, quantum chromodynamics and electron-positron annihilation
epochs (see, e.g., ref.~\cite{Carr:2024} for an excellent review and
a much more complete set of references). This possibility would have
circumvented the delay in Pop II and III star formation, allowing the
seeds to form well before decoupling at $\sim 380,000$ years. Even so,
however, it is difficult to see how this approach could have reduced
the growth time shown in figure~\ref{fig2}, which stretches the birth
event to well before the Big Bang.

\section{Conclusion}\label{conclusion}
In this {\it Letter}, we have highlighted the cosmic timeline implied
by the most recently discovered high-$z$ quasars. The reality, however,
is that a severe time compression problem with the standard model is
suggested by several diverse kinds of source, not just the very early
appearance of supermassive black holes. The so-called impossibly early
galaxy problem, first identified with the {\it Hubble Space Telescope}
\cite{Oesch:2016,Melia:2014a}, has been greatly exacerbated by other recent
{\it James Webb Space Telescope} discoveries of galaxies 
with spectroscopically confirmed redshifts up to $z\sim 14.32$ and 
several others with (less reliable) photometric redshifts up to $z\sim 16-17$ 
\cite{Pontoppidan:2022,Finkelstein:2022,Treu:2022}. The most distant
confirmed galaxies include JADES-GS-z14-0 ($z=14.32$) and JADES-GS-z14-1 
($z=13.90$) \cite{Carniani:2024} and JADES-GS-z13-0 ($z=13.2$)
\cite{Hainline:2024}. In the context of $\Lambda$CDM, the ages at which
these galaxies appeared fully formed are thus $t\sim 277$ Myr at $z\sim 14$
and $\sim 217$ Myr at $z\sim 17$.  

In the context of the standard model, these $\sim 10^9\;M_\odot$ structures
would have had to form just at the time when the first Population III stars 
were being assembled. But this situation is just as infeasible as the early 
formation and growth of black holes that we have discussed in this {\it Letter}. 
These two major problems with the standard model are, of course, closely related. 
The astrophysical principles behind the emergence of both classes of source are 
well understood, and very little, if any, room remains for the adjustments 
required to mitigate this tension.

The common thread that underlies both anomalies is the apparently incorrect
time versus redshift relation predicted by $\Lambda$CDM. Not suprisingly,
the impossibly early galaxy problem is mitigated by adopting
$R_{\rm h}=ct$ as the background cosmology \cite{Melia:2023}. These results
have signifiant implications because they rely on a period in the Universe's
expansion ($t\lesssim 2$ Gyr) when these two models differ the most. It thus
appears that all of the major issues associated with the formation of structure
in the early Universe are resolved by the inclusion into $\Lambda$CDM of the
zero active mass condition, $\rho+3p=0$, which would turn it into $R_{\rm h}=ct$
\cite{Melia:2020}.

\vskip 0.2in
\noindent{\bf Data Availability Statement:} This manuscript does not
have any associated data.

\vskip 0.2in
\noindent{\bf Code Availability Statement:} This manuscript does not
have any associated code/software.

{\acknowledgement
I am very grateful to the anonymous referee for excellent suggestions
to improve the presentation in this manuscript.
\endacknowledgement}

\bibliographystyle{JHEP3}
\bibliography{ms}

\end{document}